\documentclass[prl,twocolumn]{revtex4}
\usepackage{graphicx,subfigure}% Include figure files
\usepackage{dcolumn}% Align table columns on decimal point
\usepackage{bm}% bold math
\usepackage{amsmath}

\begin{document}
\title{\bf{VENUS ATMOSPHERE
PROFILE FROM A MAXIMUM ENTROPY PRINCIPLE}}
\author{\bf{Luis N. Epele}\thanks{E-mail: epele@fisica.unlp.edu.ar},\bf{ Huner
Fanchiotti, Carlos A. Garc\'{\i}a Canal}
\\ Laboratorio de F\'{\i}sica Te\'{o}rica,
Departamento de F\'{\i}sica, IFLP \\ Facultad de Ciencias Exactas,
Universidad Nacional de La Plata
\\C.C. 67, 1900 La Plata, Argentina.\\
\bf{Amalio F. Pacheco}\thanks{E-mail: amalio@unizar.es}\\
Facultad de Ciencias and BIFI, Universidad de Zaragoza\\ 50009
Zaragoza, Spain.\\ \bf {Jaime Sa\~nudo}\\Departamento de
F\'{\i}sica, Universidad de Extremadura\\ Badajoz, Spain. }
%\date{}
 
 %\vspace{3cm}

% \noindent AGU Index Terms: 0343; 5405; 6295
%\pagebreak
\begin{abstract}

The variational method with constraints recently developed by
Verkley and Gerkema to describe maximum-entropy atmospheric profiles
is generalized to ideal gases
 but with temperature-dependent specific heats. In so doing, an extended
 and non standard  potential
 temperature is introduced that is well suited for tackling the problem
  under consideration. This new formalism is successfully
 applied to the atmosphere of Venus. Three well defined regions emerge in this
 atmosphere up to a height of $100\, km$ from the surface: the lowest one up to
 about $35\, km$  is adiabatic, a transition layer located at the height of the
 cloud deck  and finally a third region which  is practically isothermal.

%\par
\end{abstract}
\maketitle
%\eject
\pagebreak
\section*{1. Introduction}

Since the late 1950s, there have been hundreds of space
exploration missions that have provided data on the composition,
structure and circulation of planetary atmospheres. In particular,
the vertical pressure-temperature profile of the atmosphere of
Venus was accurately observed in the Pioneer-Venus II mission in
1978 \cite{Seiff}. In ref. \cite{Landis},
an averaged data of temperature, pressure and density of the Venus
atmosphere as a function of altitude above the surface is given.
In this paper we will use this information.

Venus is the nearest planet to the Sun. The incoming solar energy,
of the order of $2600\,W/m^2$, is almost twice as great as on
Earth. Venus possess an atmosphere, which is about a hundred times
as massive as that of the Earth and its surface temperature
reaches $730\, K$. Carbon dioxide is the major component (96.5 per
cent), and 3.5 per cent of $N_2$ is the next most abundant
species.

 In a recent paper by Verkley and Gerkema \cite{Verkley},
henceforth denoted by VG, have proposed a first-principles
variational method to generate a $p-T$ profile for an atmospheric
column. This profile is the result of maximizing the entropy of
the column, $S$, maintaining fixed its mass, $M$, enthalpy, $H$,
and the integral of the potential temperature, $L$. This method is
flexible enough to describe from an isothermal profile to an
adiabatic one depending on the value of a unique parameter. VG
applied these ideas to the Earth`s troposphere and accurately
reproduced the U.S. Standard Atmosphere.

In their analysis, VG considered that the atmospheric gas is
ideal, i.e.  a gas that verifies the corresponding equation of
state, and besides, the specific heats, $C_{V,p}$, are constant.
When this second assumption is also fulfilled it is said that the
gas is perfect. These two hypotheses are quite reasonable when
dealing with the atmosphere of the Earth. For representative
temperatures and pressures of the low Venus atmosphere, however,
the temperature dependence of the $C_{V,p}$ has to inescapably be
taken into account. As our goal in this paper is the application
of the VG ideas to the Venus atmosphere, we have had to extend
their original variational method to take into account this
dependence of the specific heats on temperature. Consequently we
will first extend the VG method to ideal imperfect gases. As is
well known, the temperature dependence of the $C_{V,p}$ comes from
the excitation of some internal vibrational levels of the molecule
when the temperature is high enough.

Throughout the paper we will consider that $CO_2$ is the main
chemical component of the Venus atmosphere. The chemical
impurities coming from the $N_2$ produce negligible corrections in
our analysis. Besides, we will not consider any correction coming
from the interaction between the molecules, that is, we will stick
to the scenario of an ideal gas but with the afore-mentioned
temperature dependence of the $C_{V,p}$. Corrections of real gas
in the Venus atmosphere have been considered \cite{Staley} but
for our purposes here they are negligible. Moreover, we will
consider a constant gravity for the entire column under study
because its variation with altitude has no impact in the results.
The verification that the equation of state of ideal gases is good
to describe the gas of Venus can be done using the data of ref. \cite{Landis}. 
There one can check, at each height, that using
pressure and temperature as input, the value of density obtained
from the equation of the ideal gases is, in all cases, equal to
the experimental value with a discrepancy lower than $2$ per cent.

The index of the paper is as follows. In Section 2, we will recall
the meaning of the potential temperature, $\theta$, for ideal
perfect gases and its relation to the entropy. In Section 3, we will
review the VG method and comment on the main results when it is
applied to the Earth`s troposphere.  In Section 4, we will put in
evidence the necessity of considering an explicit dependence of the
$C_{V,p}$ with temperature for the $CO_2$ molecule, to properly
describe the Venus atmosphere. In Section 5, we will calculate the
equation of the potential temperature, i.e. the form of an
adiabatic, when dealing with ideal-imperfect gases. In Section 6, we
will extend the VG method to ideal-imperfect gases. In Section 7,
this new method will be applied to the Venus atmosphere and the
results will be commented on. Finally, in Section 8 we present our
conclusions.

\section*{2. The Potential Temperature and the Entropy
for Ideal Perfect Gases}

The first principle of Thermodynamics says that
\begin{equation}
\delta U = \delta Q - p\,\delta V.
\end{equation}
Throughout this paper, $\delta$ indicates all the variations of
thermodynamical quantities, whereas $d$ will be used to refer to
variations of quantities at different altitudes.

 For ideal gases, the equation of state, for $1\, mol$,
is
\begin{equation}
p\,V = R^{\ast}\,T
\end{equation}
(where $R^{\ast}=8.3143\, J\, mol^{-1}\, K^{-1}$ is the universal
gas constant) and the specific internal energy (internal energy
per unit mass, noted with lower case letter) reads
\begin{equation}
u = C_V\,T + constant.
\end{equation}
In an adiabatic process, $\delta Q=0$, using the first two
equations and remembering that $C_p = C_V + R^{\ast}$, we find
\begin{equation}\label{CR}
C_p\,\frac{\delta T}{T} = R^{\ast}\,\frac{\delta p}{p}.
\end{equation}

Thus, two points $(p,T)$ and $(p_r,T_r)$ belonging to the same
adiabatic trajectory fulfil
\begin{equation} \label{CRI}
C_p\,\int_{T_r}^T \frac{\delta T^{\prime}}{T^{\prime}} =
R^{\ast}\, \int_{p_r}^p \frac{\delta p^{\prime}}{p^{\prime}}.
\end{equation}
In this case, the reference temperature $T_r$ becomes the
so called potential temperature $\theta$ corresponding to 
the reference pressure $p_r$.
Note that in the previous equations, $C_p$ has been 
considered as a temperature independent 
constant because in this section we are dealing with an
ideal-perfect gas. This last equation constitutes one of the forms
of expressing Poissons equation for an adiabatic process.
Explicitly it reads as
\begin{equation}
T\,p^{-\kappa} = T_r\,p_r^{-\kappa}\,\,\,;\,\,\,\kappa =
R^{\ast}/C_p.
\end{equation}
As it is standard, choosing the reference pressure $p_r=1000\,mb$, 
the potential temperature $\theta$ reads
\begin{equation} \label{theta}
\theta = T\,\left(\frac{1000\,mb}{p} \right)^{\kappa}.
\end{equation}
The potential temperature of a gas is therefore the temperature
that it would take if we compress or expand it adiabatically to a
reference pressure of $1000\, mb$ \cite{Tsonis}. 
Obviously, the choice $p_r=1000\,mb$ is only a question of
convenience when we are dealing with the low atmosphere of the
Earth. In other words, the value of the reference pressure $p_r$
is arbitrary.

The differential expression for the specific entropy of ideal
gases is
\begin{equation} \label{SGI}
\delta s = C_p\,\frac{\delta T}{T} - R^{\ast}\,\frac{\delta p}{p}.
\end{equation}
Because of the definition of $\theta$ in Eq.(\ref{theta}), $\delta
\ln \theta = \delta \ln T - \kappa\,\delta \ln p$  and
consequently
\begin{equation} \label{SCP}
\delta s = C_p \, \delta \ln \theta,
\end{equation}
and because we are considering that $C_p$ is constant
\begin{equation} \label{SLT}
s = C_p\,\ln \theta + constant.
\end{equation}
Due to this direct relation, the potential temperature $\theta$ is
usually considered as the entropy of the gas in the physics of the
atmosphere. It is apparent that in a process of constant $\theta$,
the entropy is also constant.

\section*{3. The Verkley-Gerkema Method}

VG have proposed a first-principles variational method to generate
a $p-T$ profile for an atmospheric layer. In this method, one
considers a column of dry air with unit horizontal area in
hydrostatic equilibrium, bounded by two fixed values of the
pressure. At each altitude $z$, the local thermodynamics is
related to $p(z)$, $T(z)$ and $\rho(z)$ that can be inverted into
$z=z(p)$, $T(p)$ and $\rho(p)$, since $dp/dz = - g\,\rho$. The
profile one is looking for comes out after the maximization of the
entropy with the condition that the enthalpy and the vertical
integral of the potential temperature, $L$, are kept fixed. As the
top and bottom pressures, $p_2$ and $p_1$, of the column are
specified, the total mass of air is also fixed. We denote by
\begin{eqnarray}
S & = & \frac{C_p}{g}\, \int_{p_2}^{p_1}\,\ln \theta\,dp, \nonumber
\\
M & = & \frac{1}{g}\,\int_{p_2}^{p_1}\,dp = \frac{1}{g}\,(p_1 -
p_2), \nonumber \\
H & = & \frac{C_p}{g} \,\int_{p_2}^{p_1}\,T\,dp, \nonumber \\
L & = & \frac{C_p}{g} \,\int_{p_2}^{p_1}\,\theta\,dp,
\end{eqnarray}
the total entropy, mass, enthalpy, and vertical integral of the
potential temperature \cite{Bohren} of the air
column, respectively. The requirement of $L$ fixed is appropriate
in the case of convective mixing. An analysis on this issue and
its bibliographic antecedents are presented in ref. \cite{Verkley}.

As said above, VG imposed the maximization of $S$, with $M$, $H$,
and $L$ fixed, namely
\begin{equation}
\delta S + \lambda\,\delta H + \mu\, \delta L = 0,
\end{equation}
$\lambda$ and $\mu$ being Lagrange multipliers. This means that
one should maximizes the entropy with respect to the temperature
profile of the atmosphere with the constraints imposed. The
profile variation is consequently performed along the column
through the variation of the temperature keeping the pressure
fixed at every altitude.

 From the maximization with constraints, VG deduced
\begin{equation}
\frac{1}{T} + \lambda + \mu\,\left( \frac{p_0}{p} \right)^{\kappa}
= 0.
\end{equation}
Introducing
\[
\alpha = \frac{\mu}{\lambda},
 \]
 and denoting the temperature at
$p_0$ by $T_0$, the resulting $p-T$ profile is
\begin{equation}
T = T_0\,\frac{1 + \alpha}{1 + \alpha\,(p_0/p)^{\kappa}}.
\end{equation}

 This result is actually powerful; note that it is able to
describe a wide range of profiles, from an isothermal profile, in
the $\alpha \rightarrow 0$ limit, to an adiabatic profile in the
limit of $\alpha \rightarrow \infty$. In this last case, $p_0$ and $T_0$
clearly become $p_r$ and $\theta$ respectively. 
In the application of this
result to a specific air layer, $\alpha$ is fixed by requiring
that the ratio $H/L$, that derives from the profile, is equal to
its real value in the air layer. Then, $T_0$ is fixed by requiring
that $H$ (or $L$) is equal to its real value as well. Using this
method VG obtained a profile for the troposphere that fitted
remarkably well the U.S. Standard Atmosphere data. 
The same method has been applied to the Earth`s
mesosphere \cite{Pacheco}.

\section*{4. Specific Heat of $CO_2$}

The molecule of $CO_2$ is linear and consists of a carbon atom
that is doubly bonded to two oxygen atoms. Its molecular mass and
corresponding specific gas constant are
\[
M_{CO_2} = 44.01 g\,mol^{-1}\,\,\,\,;\,\,\,\, R = M_{CO_2}*R^{\ast} =
188.92\,J\,kg^{-1}\,K^{-1}.
\]

The various contributions to the specific heat, at constant
pressure, of a gas constituted by molecules of this type are
expressed in the form:
\begin{equation} \label{CP}
C_p = \frac{7}{2}\,R + \sum_{T_{\nu}}\,R
\,\left(\frac{T_{\nu}}{2\,T} \right)^2\,\sinh^{-2}
\left(\frac{T_{\nu}}{2\,T} \right).
\end{equation}
The first term in this sum corresponds to translation and to
rotational modes while the second term is due to vibrational
modes. The sum in the second term extends to the various
characteristic temperatures, $T_{\nu}$ , of the vibrational modes.
For $CO_2$, these values are \cite{Callen}
\begin{equation}\label{TN}
T_{\nu}(K)= 960,\, 960,\, 2000,\, 3380.
\end{equation}

\begin{figure}[bt]
\begin{center}
\includegraphics[height=8cm,angle=270]{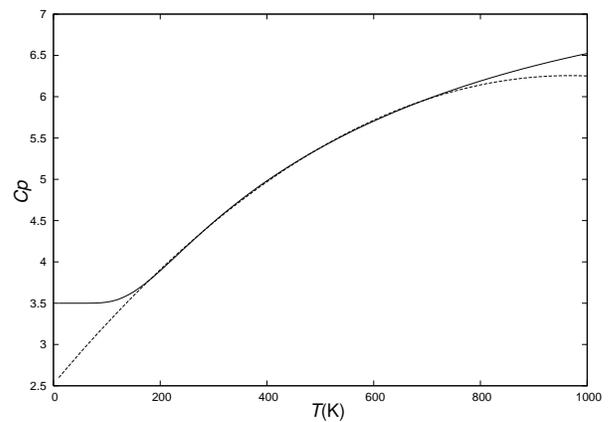}
\end{center}
\caption{Specific heat at constant pressure ($C_p$) in units
of $R^{\ast}$ as a function of temperature ($T$), of the $CO_2$
molecule (continuous line) and its polynomial fit (dashed line). For
details see Section 4}
\label{figure-open}
\end{figure}

For practical uses, it is convenient to express  Eq.(\ref{CP}) by a
simple algebraic approximation. Thus, it is standard to express the
 specific heat by a series of powers in $T$, with coefficients empirically adjusted.
 Three terms, namely
\begin{equation}\label{CC}
C_p= A + B\,T + C\,T^2,
\end{equation}
 are enough to describe the specific heat with an excellent accuracy
 in the range we are interested in.
This is clearly shown in Fig.1, where we have plotted $C_p$ vs. $T$
for $CO_2$. The continuous line corresponds to the ''exact" values
which have been obtained by plotting (\ref{CP}) with the insertion
of the four characteristic temperatures of vibration as given in
(\ref{TN}). The dotted line is the polynomial approximation
(\ref{CC}), the coefficients being
\[
 A= 2.5223,\,\,\,\, B= 0.77101\, 10^{-2},\,\,\,\, C=
-0.3981\,10^{-5}.
\]
 The units
of $A$, $B$ and $C$ are $R^{\ast}$, $R^{\ast}/K$ and $R^{\ast}/K^2$
respectively. From Fig.1 it is apparent, first, that $C_p$ has a
significant change in the range of temperatures of the Venus
atmosphere, from $735\, K$ at $z=0$ to $175\, K$ at $z=100\, km$.
Even if we restrict our analysis to the lowest layer, i.e., from
$z=0$ to $z=40\, km$, the interval of temperature is between $735\,
K$ and $385\, K$. In Fig.1 one first observes that the change in
$C_p$ in this interval of temperature is of the order of  $50 \%$,
and secondly, that this polynomial fit is excellent. Thus, we will
use (\ref{CC}) as input to generalize the VG method and apply it to
the atmosphere of Venus.

\section*{5.  Potential Temperature for a Temperature-Dependent
Specific Heat}

The generalization of the VG method can start from Eq.(\ref{CR})
because it is applicable to any ideal gas. In this Section
however, Eq.(\ref{CRI}) for the adiabatic transformations, adopts the form
\begin{equation}
\int_{\theta}^T\,C_p(T^{\prime})\,\frac{\delta
T^{\prime}}{T^{\prime}} \equiv \int_{\theta}^T\, (A + B\,T^{\prime} +
C\,T^{\prime \, 2})\,\frac{\delta T^{\prime}}{T^{\prime}} =
R^{\ast}\,\int_{p_r}^p\,\frac{\delta p^{\prime}}{p^{\prime}}.
\end{equation}
$p_r$ being the pressure level of reference and $\theta$ the potential
temperature characteristic of this adiabatic trajectory. We find
\begin{equation}\label{thetabar}
\frac{p}{p_r} = \left(\frac{T}{\theta}
\right)^{A/R^{\ast}}\,\exp\left[\frac{B\,(T-\theta)}{R^{\ast}}\right]\,
\exp\left[\frac{C\,(T^2-{\theta}^2)}{2\,R^{\ast}}\right].
\end{equation}

\begin{figure}[ht]
\begin{center}
\includegraphics[width=8cm]{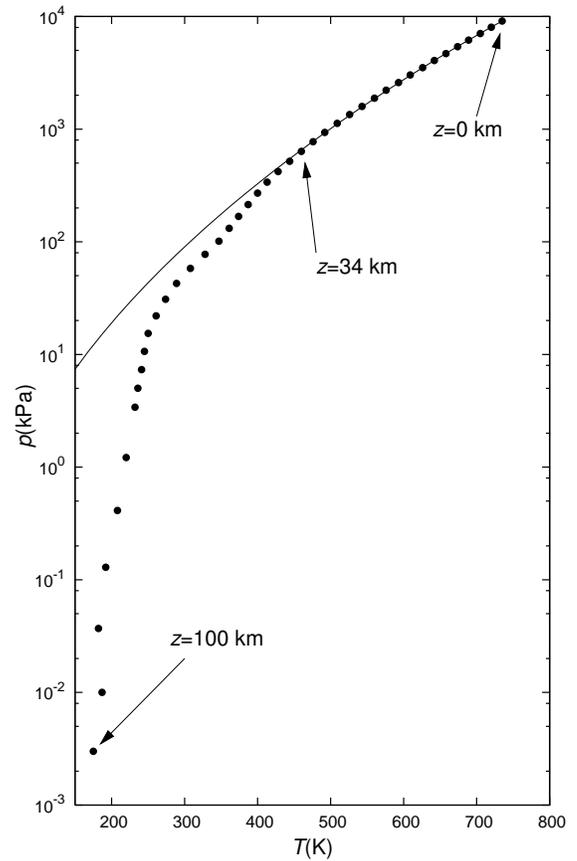}
\end{center}
\caption{ Pressure-temperature profile of the atmosphere of
Venus. The data (black dots) have been taken from Table 1, and the
fitting line is an adiabatic as given by Eq.(\ref{thetabar}).
}
\label{figure2}
\end{figure}

In Fig.2, we have adjusted the data of the low atmosphere of Venus
by means of this formula for an adiabatic. There is a generalized
consensus supporting this behavior as can be read in ref. \cite{Seiff}.
 It shows an excellent agreement along the lowest $30\, km$
and a clear departure for upper heights.

In the present case, the physical meaning of the potential 
temperature is the same as
that of (\ref{theta}), i.e. the temperature that the gas would take
if we compressed or expanded it adiabatically to a pressure $p_r$.
The price to pay now is that given the coordinates $(p,T)$, the
value of $\theta$ has to be computed numerically by
iterations. Of course, in the case $B=C=0$, (\ref{theta}) and
(\ref{thetabar}) are the same equation.

The differential expression for the specific entropy of ideal
gases was written in (\ref{SGI}) and now reads
\begin{equation} \label{SCPT}
\delta s = C_p(T)\,\frac{\delta T}{T} - R^{\ast}\,\frac{\delta
p}{p}.
\end{equation}

For practical purposes, let us define a new magnitude $\tau$, with physical dimensions of
temperature, in the following way:
\begin{equation} \label{CPT}
C_p(T)\,\frac{\delta T}{T} = C_p^0\,\frac{\delta \tau}{\tau},
\end{equation}
with
\begin{equation}
C_p^0 = C_p(T_0)\,\,\,;\,\,\,\tau(T_0) = T_0.
\end{equation}

In fact, this new variable $\tau$,that reduces 
to the standard temperature as soon as  $C_p$ is constant, 
allows one to treat this problem in exactly the same way as
in the case of the ideal perfect gas.
 
After specifying $C_p(T)$ in a form like (\ref{CC}), and by the
integration of (\ref{CPT}) we obtain
\begin{eqnarray}
C_p^0\,\int_{T_0}^{\tau}\,\frac{\delta \tau}{\tau} & = &
\int_{T_0}^{T}\, \left(A + B\, T + C\, T^2 \right)\,\frac{\delta
T}{T}
\nonumber \\
& = & A\,\left[\ln \left(\frac{T}{T_0} \right) +
\frac{B}{A}\,\left(T - T_0\right) + \frac{C}{2\,A}\,\left( T^2 -
T_0^2 \right)\right],
\end{eqnarray}
 and the result is
\begin{equation} \label{LTT}
\ln\left(\frac{\tau}{T_0}\right) =
\frac{A}{C_p^0}\,\ln\left\{\frac{T}{T_0}\,\exp\left[\frac{B}{A}\,\left(T
- T_0\right) + \frac{C}{2\,A}\,\left( T^2 - T_0^2
\right)\right]\right\}.
\end{equation}

This is the relation between $\tau$ and $T$. This formula fulfils
the two agreed upon conditions: $\tau(T_0) = T_0$ and $\tau(T) =
T$ for $B=C=0$

Using (\ref{CPT}) in (\ref{SCPT})
\begin{equation}
\delta s = C_p^0\,\left( \frac{\delta \tau}{\tau} -
\frac{R^{\ast}}{C_p^0}\,\frac{\delta p}{p} \right),
\end{equation}
and in a similar way to what was done to arrive to (\ref{SLT}), we
obtain
\begin{equation} \label{STB}
s = C_p^0\,\ln\tilde{\tau} + constant,
\end{equation}
with
\begin{equation} \label{TTT}
\tilde{\tau} = \tau\,\left(\frac{p}{p_r}\right)^{-\kappa^0},
\end{equation}
and
\begin{equation} \kappa^0 =
\frac{R^{\ast}}{C_p^0}.
\end{equation}
 It is now clear from this result that in an isentropic process, 
 $\tilde{\tau}$ is also conserved. 
The use of (\ref{STB}) will
permit us to extend the VG method quite easily, at least from a
conceptual point of view.

\section*{6. Generalization of the VG Method}

 The formal analogy between these new magnitudes and those
appearing in Section 2, $\tilde{\tau} \leftrightarrow \theta$,
$\kappa^0 \leftrightarrow \kappa$  , permits us to extend the VG
method reviewed in Section 3 in a quite natural way. The entropy,
the enthalpy and the L function of an atmospheric column per unit
section are:
\begin{equation}
S = \frac{C_p^0}{g}\, \int_{p_2}^{p_1}\,\ln\tilde{\tau}\,dp,
\end{equation}
\begin{equation}
H = \frac{1}{g}\,\int_{p_2}^{p_1}\,h\,dp = \frac{1}{g}\,
\int_{p_2}^{p_1}\,C_p(T)\,T\,dp,
\end{equation}
\begin{equation}
L = \frac{C_p^0}{g}\, \int_{p_2}^{p_1}\,\tilde{\tau}\,dp.
\end{equation}

 The condition of maximum $S$, with $H$ and $L$ constants, is expressed
as the extreme of the functional $\Psi$ under the variation of the
profile $T(p)$, in the following way
\begin{equation} \label{DPSI}
\delta \Psi =\int_{p_2}^{p_1}\,\frac{C_p^0}{g}\, \int_{p_2}^{p_1}\,
\delta\left[\ln\tilde{\tau} + \mu\,\tilde{\tau}\right]\,dp +
\lambda\,\frac{1}{g}\,\int_{p_2}^{p_1}\,\delta[h]\,dp = 0.
\end{equation}

\begin{figure}[ht]
\begin{center}
\includegraphics[width=8cm]{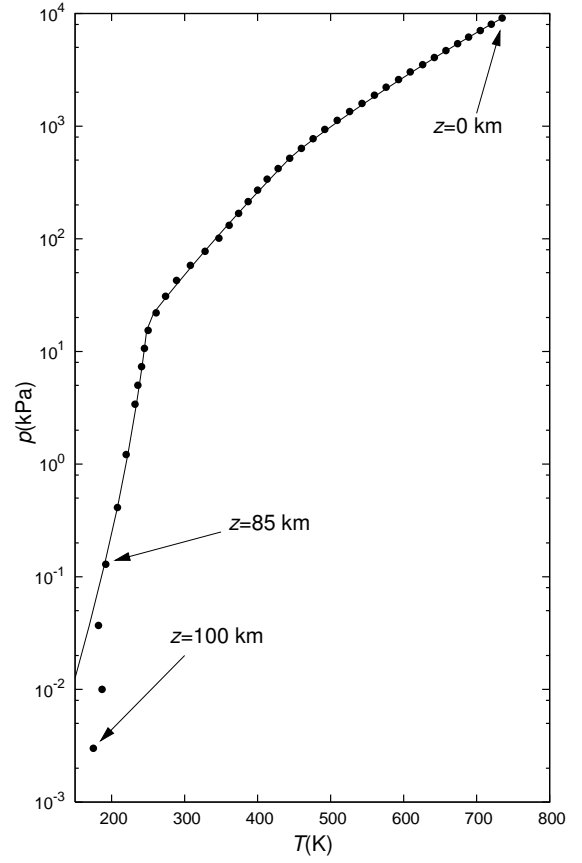}
\end{center}
\caption{ Pressure-temperature profile of the atmosphere of
Venus. The data (black dots) come from Table 1 and the theoretical
results (continuous line) have been obtained using the extended VG
method.}
\label{figure3}
\end{figure}

 To go into the details of this variational problem, remember
that
\begin{equation}
\tau =\tau(T)\,\,\,;\,\,\,h=h(T)\,\,\,;\,\,\,\tilde{\tau} =
\tilde{\tau}(\tau,p),
\end{equation}
 Then, the
condition $\delta \Psi = 0$ ($p$ fixed) implies that
\begin{equation}
\left(\frac{1}{\tilde{\tau}} + \mu \right)\,\frac{\delta
\tilde{\tau}}{\delta T} + \frac{\lambda}{C_p^0}\,\frac{\delta
h}{\delta T} =0,
\end{equation}
and taking into account that
\begin{equation}
T\,C_p^0\,\frac{\delta \ln\tilde{\tau}}{\delta T}  =
\frac{\delta h}{\delta T},
\end{equation}
one gets
\begin{equation}
\left(\frac{1}{T} + \mu \,\frac{\tilde{\tau}}{T}\right) + \lambda
=0.
\end{equation}

Finally, the result of the extreme can be written as
\begin{equation}\label{RES}
T = T_0\,\frac{1 + \alpha}{1 + \alpha\,\tilde{\tau}/T},
\end{equation}
with
\begin{equation}
\alpha = \frac{\mu}{\lambda}\,\,\,\,\,;\,\,\,\,\,T_0
=\frac{1}{\lambda},
\end{equation}
being formally identical to the VG result. Nevertheless, it does
not provide an explicit expression of $T$ as a function of $p$.
This is due to the fact that $\tilde{\tau}$ depends on $\tau$
and through it on $T$, so that $\tilde{\tau}/T$ is now a
function of $p$ and $T$. Consequently, the equation (\ref{RES})
has to be treated numerically. In fact, by using the method of
Lagrange multipliers, the experimental data fix the constraints
$H$ and $L$. Then, the parameters $T_0$ and $\alpha$ are
determined in order to satisfy $H$ and $L$.

\section*{7. Results of the Extended VG Method}

The performance of the extended VG method, summarized in Eq.
(\ref{RES}), when applied to the atmosphere of Venus is shown in
Fig.3. Note that the logarithmic scale used for pressure is more
sensitive and hence puts more clearly in evidence the virtues or
flaws of the fit. The characteristics of the Venus atmosphere
allow us to well distinguish three regions in altitude. In each
one of these regions we have applied the extended VG method by
maximizing the entropy with both $H$ and $L$ constraints.

In the lowest one, from $z=0$ to $z\simeq 34\, km$, we found that
the value of $\alpha$ is $44$ and the value of $T_0 = 735 \, K$.
This high value of the unique parameter of the model indicates the
clear preponderant role of $L$ with respect to $H$, as the
dominant constraint in the maximization of $S$. From this fact we
conclude that in this layer the profile is adiabatic. This clearly
agrees with the result shown in Fig.2, obtained  by means of
Eq.(\ref{thetabar}).

In the layer from $z\simeq 60\, km$ to $z\simeq 100\, km$, the
value obtained for $\alpha$ is $0.18$ and $T_0 = 245\, K$ This
indicates that here $H$ is the dominant constraint, and in
consequence the profile of this layer is basically isothermal.

The method assigns $\alpha =3$ with $T_0 = 444 \, K$ to the
intermediate layer, a transition region, from $z\simeq 34\,Km$ to
$z\simeq 60\, Km$.

\section*{8. Discussion and Conclusions}

Our purpose in this paper was to extend the first-principles VG
method to atmospheric layers where the thermal deviation is so
high that it is not reasonably possible to consider a constant
unique value for the specific heat in all the points of the layer.
To implement this extension, we have had to define a new extended
potential temperature, $\tilde{\tau}$. A potential temperature
is, by definition, a generalized temperature that characterizes
adiabatic processes, or in other words, that is conserved in
processes where the entropy is constant. The relation between the
new $\tilde{\tau}$ and entropy maintains the same standard form
as that existing for ideal-perfect gases.

>From the surface of Venus up to about $100\, km$, we have
distinguished $3$ layers. The extended VG method, for each layer,
after fixing $T_0$ to the corresponding initial temperature of the
layer, deals with a unique parameter $\alpha$ to adjust the
constraints. From the surface up to $35\, km$, the method assigns
a high value of $\alpha$. This implies that the $p-T$ profile here
is adiabatic. There is no surprise in this result; an adiabatic
lapse-rate was detected in the low atmosphere of Venus during the
first observations of the Russian Venera probes. This was verified
later in other spatial missions \cite{Seiff}. In the
layer from $35$ to $65\, km$ in height $\alpha$ has an
intermediate value that indicates a transition. Finally, from $65$
to about $100\, km$, this method produces a low $\alpha$, which is
interpreted as an isothermal layer. This is the first time that
the VG method identifies an adiabatic layer where it should; this
concordance shows clearly the success of the method itself and of
the extension presented in this paper.

\section*{Acknowledgments}

LNE, HF and CAGC acknowledge CONICET and ANPCyT of Argentina for
financial support. AFP and JS thank the Spanish DGICYT for
financial support (Project FIS 2005-06237).

%\section*{References}


\begin{thebibliography}{K}

\bibitem{Seiff}
\noindent A. Seiff et al, 
J. Geophys. Res. {\bf 85}, 7903 (1980).

\bibitem{Landis}
\noindent  G. Landis, A. Colozza and C. LaMarre,
{\it Atmospheric flight on Venus}, (NASA-2002-0919, 2002).

\bibitem{Verkley}
\noindent W.T.M. Verkley and T. Gerkema, 
J. of the Atmospheric Sciences {\bf 61}, 931 (2004).

\bibitem{Staley}
\noindent D.O. Staley, 
J. Atmos. Sci. {\bf 27}, 219 (1970).

\bibitem{Tsonis}
\noindent A.A. Tsonis, {\it An Introduction to Atmospheric 
Thermodynamics}, (Cambridge Univ. Press, 2002).

\bibitem{Bohren}
C.F. Bohren and B.A. Albrecht, {\it Atmospheric
Thermodynamics}, (Oxford Univ. Press, 1998).

\bibitem{Pacheco}
\noindent A.F. Pacheco and J. Sa\~nudo, 
Nuovo Cimento {\bf C28}, 29 (2005).

\bibitem{Callen}
\noindent H.B. Callen, {\it Thermodynamics}, (John-Wiley and
Sons, 1980).

\end{thebibliography}
\end{document}